\documentclass[a4paper]{jpconf} 
\usepackage{graphicx}
\begin{document}
\title{A novel theoretical approach for the study of resonances in weakly bound systems}

\author{Mahamadun Hasan$^{1}$, Shamim Haque Mondal$^{1}$, Murshid Alam$^{1}$, Tasrief Surungan$^{2}$ and Md Abdul Khan$^{1^*}$}

\address{$^{1}$Department of Physics, Aliah University, IIA/27, Newtown, Kolkata-700160, India}
\address{$^{2}$Department of Physics, Hasanuddin University, Makassar-90245, Indonesia}
\ead{$^{*}$drakhan.rsm.phys@gmail.com;drakhan.phys@aliah.ac.in ($^*$Corresponding author)}

\begin{abstract}
In this paper, a novel theoretical scheme is presented to investigate resonant levels
in weakly bound nuclear systems by the use of isospectral potentials. In this scheme, a new
potential is constructed which is strictly isospectral with the original shallow-well potential and
has properties that are desirable to calculate resonances more accurately and
easier. Effectiveness of the method has been had been demonstrated in terms of its application to the first $0^+$ resonances in the neutron-rich isotopes $^A$C ($\equiv ^{A-2}$C+n+n) in the three-body cluster model for A=18, 20.
\end{abstract}
\vspace{5pt}17
PACS number(s): 21.45.+v, 21.10.Dr, 27.10.+h \\
Keywords: Halo nuclei, resonance, isospectral potential.

\section{Introduction}
The discovery of halo nuclei (neutron-halo and proton-halo) in the neighborhood of drip lines is one of the major achievements of the advancements of the radioactive ion-beam facilities. Halo structure is characterized by a relatively stable and denser core surrounded by weakly bound one or more valence nucleon(s) giving rise to long extended tail in the density distribution. This low-density tail is supposed to be the consequence of quantum mechanical tunneling of the last nucleon(s) through a-shallow barrier following an attractive well that appears due to the short-range nuclear interaction, at energies smaller than the height of the barrier. In halo nuclei, one seldom finds any excited bound states because of the utmost support one bound state at energies less than 1 MeV. Halo nuclei have high scientific significance as they exhibit one or more resonance state(s) just above the binding threshold. The observed halo nuclei-$^{17}$B, $^{19}$C show one-neutron halo; $^6$He, $^{11}$Li, $^{11, 14}$Be show two-neutron halo; $^8$B, $^{26}$P show one proton-halo; $^{17}$Ne, $^{27}$S show two proton-halo and $^{14}$Be, $^{19}$B show four neutron-halo structure respectively \cite{tanihata-1985, kobayashi-2012, hwash-2017, jensen-2000, schwab-2000, tanaka-2010}. Halo-nuclei is characterized by their unusually large r.m.s. matter radii (larger than the liquid-drop model prediction of $R_A\propto A^{1/3}$) \cite{audi-2003, acharya-2013} and sufficiently small two-nucleon separation energies (typically less than 1 MeV). Tanaka et al. 2010. \cite{tanaka-2010} observed of a large reaction cross-section in the drip-line nucleus $^{22}$C, Kobayashi et al. 2012 \cite{kobayashi-2012}, conducted research on one- and two-neutron removal reactions from the most neutron-rich carbon isotopes, Gaudefroy et al 2012 \cite{gaudefroy-2012} carried a direct mass measurements of $^{19}$B, $^{22}$C, $^{29}$F, $^{31}$Ne, $^{34}$Na and some other light exotic nuclei. Togano et al, 2016 \cite{togano-2016} studied interaction cross-section of the two-neutron halo nucleus $^{22}$C. \newline Nuclear matter distribution profile of such nuclei has an extended low-density tail forming a halo around the more localized dense nuclear core. Thus, in addition to bound state properties, continuum spectra is another significant parameter that is highly involved in the investigation of structure and interparticle interactions in the exotic few-body systems like the halo nuclei. It is worth stating here that the study of resonances is of particular interest in many branches of physics involving weakly bound systems in which only few bound states are possible.\newline In the literature survey, we found three main theoretical approaches that were used to explore the structure of 2n-halo nuclei. The first one is the microscopic model approach in which the valence neutrons are supposed to move around the conglomerate of other nucleons (protons and neutrons) without having any stable core. The second one is the three-body cluster model in which the valence nucleons are assumed to move around the structureless inert core. And the third one is the microscopic cluster model in which the valence nucleons move around the deformed excited core \cite{saaf-2014, nesterov-2010, korennov-2004}. There are several theoretical approaches which are employed for computation of resonant states. Some of those are the positive energy solution of the Faddeev equation \cite{cobis-1997}, complex coordinate rotation (CCR) \cite{csoto-1993, ayoma-1995}, the analytic computation of bound state energies \cite{tanaka-1997}, the algebraic version of resonating group method (RGM) \cite{vasilevsky-2001}, continuum-discretized coupled-channels (CDCC) method clubbed to the cluster-orbital shell model (COSM) \cite{ogata-2013}, hyperspherical harmonics method (HHM) for scattering states \cite{danilin-1997}, etc. In most of the theoretical approaches, Jacobi coordinates are used to derive the relative coordinates separating the center of mass motion.
\newline One of the most challenging obstacles that are involved in the calculation of resonances in any weakly bound nucleus is the large degree of computational error. In our case, we overcome this obstacle by adopting a novel theoretical approach by interfacing the algebra of supersymmetric quantum mechanics with the algebra involved in the hyperspherical harmonics expansion method. In this scheme, one can handle the ground state as well as the resonant states on the same footing. The technique is based on the fact that, for any arbitrarily given potential (say, {$U$}), one can construct a family of isospectral potentials ($\hat{U}$), in which the latter depends on an adjustable parameter ($\lambda$). And when the original potential has a significantly low and excessively wide barrier (poorly supporting the resonant state), $\lambda$ can be chosen judiciously to enhance the depth of the well together with the height of the barrier in $\hat{U}$. This enhanced well-barrier combination in $\hat{U}$ facilitates trapping of the particle which in turn facilitates the computation of resonant state more accurately at the same energy, as that in the case of {$U$}. This is because, {$U$} and $\hat{U}$ are {\bf strictly isospectral}.
\newline To test the effectiveness of the scheme we apply the scheme to the first $0^+$ resonant states of the carbon isotopes $^A$C, for A equal to 18 and 20 respectively. We chose three-body (2n+$^{\rm A-2}$C) cluster model for each of the above isotopes, where outer core neutrons move around the relatively heavier core $^{A-2}$C. The lowest eigen potential derived for the three-body systems has a shallow well following a skinny and sufficiently wide barrier. This skinny-wide barrier gives rise to a large resonance width. One can, in principle, find quasi-bound states in such a shallow potential, but that poses a difficult numerical task. For a finite height of the barrier, a particle can temporarily be trapped in the shallow well when its energy is close to the resonance energy. However, there is a finite possibility that the particle may creep in and tunnel out through the barrier. Thus, a more accurate calculation of resonance energy is easily masked by the large resonance width resulting from a large tunneling probability due to a low barrier height. Hence, a straightforward calculation of the resonance energies of such systems fails to yield accurate results. 
\newline We adopt the hyperspherical harmonics expansion method (HHEM) \cite{ballot-1982} to solve the three-body Schr\"{o}dinger equation in relative coordinates. In HHEM, three-body relative wavefunction is expanded in a complete set of hyperspherical harmonics. The substitution of the wavefunction in the Schr\"{o}dinger equation and use of orthonormality of HH gives rise to an infinite set of coupled differential equations (CDE). The method is an essentially exact one, involving no other approximation except an eventual truncation of the expansion basis subject to the desired precision in the energy snd the capacity of available computer. However, hyperspherical convergence theorem \cite{schneider-1972} permits extrapolation of the data computed for the finite size of the expansion basis, to estimate those for even larger expansion bases. However, the convergence of HH expansion being significantly slow one needs to solve a large number of CDE's to achieve desired precision causing another limitation, hence we used the hyperspherical adiabatic approximation (HAA) \cite{ballot1-1982}to construct single differential equation (SDE) to be solved for the lowest eigen potential, ${U}_0(\rho)$) to get the ground state energy $E_0$ and the corresponding wavefunction $\psi_0(\rho)$ \cite{das-1982}.
\newline We next derive the isospectral potential $\hat{U}(\lambda,\rho)$ following algebra of the SSQM \cite{cooper-1995, khare-1989, nieto-1984}. Finally, we solve the SDE for $\hat{U}(\lambda,\rho)$ for various positive energies to get the wavefunction. We then compute the probability density corresponding to the wavefunction for finding the particle within the deep-sharp well following the enhanced barrier. A plot of probability density as a function of energy shows a sharp peak at the resonance energy. The actual width of resonance can be obtained by back-transforming the wave function $\hat{\psi}(\lambda, \rho)$ corresponding to $\hat{U}(\lambda,\rho)$ to $\psi(\rho)$ of $U(\rho)$. 
\newline The paper is organized as follows. In sections 2, we briefly review the HHE method. In section 3, we present a precise description of the SSQM algebra to construct the one-parameter family of isospectral potential $\hat{U}(\lambda,\rho)$. The results of our calculation are presented in section 4 while conclusions are drawn in section 5. 

\section{Hyperspherical Harmonics Expansion Method}
For a the three-body model of the nuclei $^{A-2}$C+n+n, the relatively heavy core $^{A-2}$C is labeled as particle 1, and two valence neutrons are labelled as particle 2 and 3 respectively. Thus there are three possibile partitions for the choice of Jacobi coordinates. In any chosen partition, say the $i^{th}$ partition, particle labelled $i$ plays the role of spectator while remaining two paricles form the interacting pair. In this partition the Jacobi coordinates are defined as
\begin{equation}
\vec{x_{i}} = a_i(\vec{r_{j}} - \vec{r_{k}}); \vec{y_{i}} = \frac{1}{a_i} \left(\vec{r_{i}} -\frac{m_{j}\vec{r_{j}} + m_{k} \vec{r_{k}}}{ m_{j} + m_{k}} \right); \vec{R}=  \frac{\sum_{i=1}^3 m_{i}\vec{r_{i}}}{M}      \end{equation}
where $i,j, k$ form a cyclic permutation of 1,2,3. The parameter $a_i= \left[\frac{m_{j} m_{k}M}{m_{i}(m_{j}+m_{k})^{2}} \right]^{\frac{1}{4}}$; $m_{i}, \vec{r_{i}}$ are the mass and position of the $i^{th}$ particle and $M(=\sum_{i=1}^3m_{i})$, $\vec{R}$ are those of the centre of mass (CM) of the system. Then in terms of Jacobi coordinates, the relative motion of the three-body system can be described by the equation \begin{equation}
\left\{ - \frac{\hbar^{2}}{2\mu} \nabla_{x_{i}}^{2}- \frac{\hbar^{2}}{2\mu} \nabla_{y_{i}}^{2}+ 
V_{jk} (\vec{x_{i}})
 +V_{ki} (\vec{x_{i}}, \vec{y_{i}} ) + V_{ij} (\vec{x_{i}}, \vec{y_{i}})-E
\:  \right\} \Psi (\vec{x_{i}}, \vec{y_{i}}) = 0
\end{equation}
where ${\mu = \left[ \frac{m_{i} m_{j} m_{k}}{M} \right]^{\frac{1}{2}}}\rightarrow$ is the reduced mass of the system, $V_{ij}$ represents the interaction potential between the particles $i$ and $j$, $x_{i} = \rho \cos \phi_{i}$; $y_{i}= \rho \sin \phi_{i}$; $\phi_i=\tan^{-1}(\frac{y_i}{x_i})$; $\rho=\sqrt{x_i^2+y_i^2}$. The hyperradius $\rho$ together with five angular variables $\Omega_{i} \rightarrow \{\phi_{i}, \theta_{x_{i}}, {\cal \phi}_{x_{i}}, \theta_{y_{i}}, {\cal \phi}_{y_{i}} \}$ constitute hyperspherical coordinates of the system. The Schr\"{o}dinger equation in hyperspherical variables $(\rho, \Omega_{i})$ becomes
\begin{equation}
\left\{ - \frac{\hbar^{2}}{2\mu}\frac{1}{\rho^5} \frac{\partial^2}{\partial\rho^2} - \frac{\hbar^{2}}{2\mu}\frac{4}{\rho}\frac{\partial}{\partial\rho}+
\frac{\hbar^{2}}{2\mu}\frac{\hat{{\cal K}}^{2}(\Omega_{i})}{\rho^{2}} + V(\rho, \Omega_{i}
) - E \right\} \Psi(\rho, \Omega_{i})= 0.
\end{equation}
In Eq.(3) $V(\rho, \Omega_{i}) = V_{jk} + V_{ki} + V_{ij}$ is the total interaction potential in the $i^{th}$ partition and $\sl{\hat{{\cal K}}^{2}}(\Omega_{i})$ is the square of the hyperangular momentum operator satisfying the eigenvalue equation 
\begin{equation}
\hat{{\cal K}}^{2}(\Omega_{i}) {\cal Y}_{K \alpha_{i}}(\Omega_{i}) = K (K + 4) {\cal Y}_{K \alpha_{i}}(\Omega_{i}) \end{equation}
$K$ is the hyperangular momentum quantum number and $\alpha_{i}$ $\equiv\{l_{x_{i}}, l_{y_{i}}, L, M \}$, ${\cal Y}_{K\alpha_{i}}(\Omega_{i})$ are the hyperspherical harmonics (HH) for which a closed analytic expressions can be found in ref. \cite{cobis-1997}. 
\newline In the HHEM, $\Psi(\rho, \Omega_{i})$ is expanded in the complete set of HH corresponding to the partition "$i$" as 
\begin{equation}
\Psi(\rho, \Omega_{i}) = \sum_{K\alpha_{i}}\frac{\psi_{K\alpha_{i}} (\rho)}{\rho^{5/2}} {\cal Y}_{K\alpha_{i}}(\Omega_{i}) \end{equation}
Use of Eq. (5), in Eq. (3) and application of the orthonormality of HH leads to a set of coupled differential equations (CDE) in $\rho$
\begin{equation}
\left\{-\frac{\hbar^{2}}{2\mu}\frac{d^{2}}{d\rho^{2}}
+\frac{\hbar^{2}}{2\mu}\frac{(K+3/2)(K+5/2)}{\rho^2}-E\right\} 
\psi_{K\alpha_{i}}(\rho)
+\sum_{K^{\prime}\alpha_{i}^{\prime}} {\cal M}_{K\alpha_{i}}^{K^{\prime}\alpha_{i}^{\prime}}\psi_{K^{\prime}\alpha_{i}^{\prime}}(\rho)=0.
\end{equation}
where \begin{equation}
{\cal M}_{K\alpha_{i}}^{K^{\prime}\alpha_{i}^{\prime}} = \int {\cal Y}_{K\alpha_{i}}^{*}(\Omega_{i}) V(\rho, \Omega_{i}) {\cal Y}_{K^{\prime} \alpha_{i}^{~\prime}}(\Omega_{i}) d\Omega_{i}.
\end{equation}
The infinite set of CDE's represented by Eq. (6) is truncated to a finite set by retaining all K values up to a maximum of $K_{max}$ in the expansion (5). For a given $K$, all allowed values of $\alpha_{i}$ are included. The size of the basis states is further restricted by symmetry requirements and associated conserved quantum numbers. The reduced set of CDE's are then solved by adopting hyperspherical adiabatic approximation (HAA) \cite{ballot1-1982}. In HAA, the CDE's are approximated by a single differential equation assuming that the hyperradial motion is much slower compared to hyperangular motion. For this reason, the angular part is first solved for a fixed value of $\rho$. This involves diagonalization of the potential matrix (including the hyper centrifugal repulsion term) for each $\rho$-mesh point and choosing the lowest eigenvalue $U_0(\rho)$ as the lowest eigen potential \cite{das-1982}. Then the energy of the system is obtained by solving the hyperradial motion for the chosen lowest eigen potential ($U_0(\rho)$), which is the effective potential for the hyperradial motion \begin{equation}
\left\{-\frac{\hbar^{2}}{2\mu}\frac{d^{2}}{d\rho^{2}} + U_0(\rho) - E \right\} \psi_{0}(\rho) = 0 \end{equation}
Renormalized Numerov algorithm subject to appropriate boundary 
conditions in the limit $\rho \rightarrow 0$ and $\rho\rightarrow\infty$ 
is then applied to solve Eq. (8) for E ($\leq E_0$). The hyper-partial wave $\psi_{K\alpha_{i}}(\rho)$ is given by 
\begin{equation}
\psi_{K \alpha_{i}}(\rho) = \psi_{0}(\rho) \chi_{K \alpha_{i},0}(\rho)
\end{equation}
where $\chi_{K \alpha_{i},0}(\rho)$ is the ${(K\alpha_{i})}^{th}$ element of the eigenvector, corresponding to the lowest eigen potential $U_0(\rho)$.

\section{Construction of Isospectral Potential}
In this section we present a bird's eye view of the scheme of construction of one parameter family of isospectral potentials. We have from Eq. (8) \begin{equation}
U_0(\rho) = E_0 + \frac{\hbar^{2}}{2\mu}\frac{\psi_0^{\prime\prime}(\rho)}{\psi_0(\rho)} \end{equation}
In 1-D supersymmetric quantum mechanics, one defines a superpotential for a system in terms of its ground state wave function ($\psi_{0}$) \cite{cooper-1995} as
\begin{equation}
W(\rho)= -\frac{\hbar}{\sqrt{2m}}\frac{\psi_{0}^{\prime}(\rho)}{\psi_{0}(\rho)}.
\end{equation}
The energy scale is next shifted by the ground state energy $(E_{0})$ of the 
potential $U_0(\rho)$, so that in this shifted energy scale the new potential become
\begin{equation} 
U_1(\rho) = U_0(\rho) - E_{0}=\frac{\hbar^{2}}{2\mu}\frac{\psi_0^{\prime\prime}(\rho)}{\psi_0(\rho)}\end{equation} 
having its ground state at zero energy. One can then easily verify that $U_1(\rho)$ is expressible in terms of the superpotential via the Riccati equation
\begin{equation}
U_1(\rho) = W^{2}(\rho) - \frac{\hbar}{\sqrt{2m}}W^{\prime }(\rho).
\end{equation}
By introducing the operator pairs
\begin{equation}
\left. \begin{array}{lcl}
A^{\dag}&  =&  -\frac{\hbar}{\sqrt{2m}}\frac{d}{d\rho}+W(\rho)\\
A  &=&  \frac{\hbar}{\sqrt{2m}}\frac{d}{d\rho}+W(\rho)
\end{array} \right\}
\end{equation}
the Hamiltonian for $U_1$ becomes
\begin{equation}
H_1 = -\frac{\hbar^{2}}{2m} \frac{d^{2}}{d\rho^{2}} + U_1(\rho) = A^{\dag}A.
\end{equation}
The pair of opertors $A^{\dag}, A$ serve the purpose of creation and annihilation of nodes in the wave function. Next we introduce a partner Hamiltonian $H_{2}$, corresponding to the SUSY partner potential $U_2$ of $U_1$ as
\begin{equation}
H_{2}=-\frac{\hbar^{2}}{2m}\frac{d^{2}}{d\rho^{2}}+U_2(\rho)=AA^{\dag}
\end{equation}
where 
\begin{equation}
U_2(\rho)=W^{2}(\rho)+\frac{\hbar}{\sqrt{2m}}W^{\prime}(\rho).
\end{equation}
Energy eigen values and wavefunctons corresponding to the SUSY partner Hamiltonians $H_1$ and $H_2$ are connected via the relations
\begin{equation}
\left. \begin{array}{lcl}
E_n^{(2)} & = & E_{n+1}^{(1)}, E_0^{(1)}=0 \; (n=0, 1, 2, 3,...),\\
\psi_n^{(n)}&=&\sqrt{E_{n+1}^{(1)}}A\psi_{n+1}^{(1)}\\
\psi_{n+1}^{(1)}&=&\sqrt{E_n^{(2)}}A^{\dagger}\psi_{n}^{(2)}\\
\end{array} \right\}\\
\end{equation}
where $E_{n}^{(i)}$ represents the energy of the $n^{th}$ excited state of $H_{i}$ (i=1, 2). Thus $H_{1}$ and $H_{2}$ have identical spectra, except the fact that the partner state of $H_{2}$ corresponding to the ground state of $H_{1}$ is absent in the spectrum of $H_{2}$ \cite{cooper-1995}. Hence the potentials $U_1$ and $U_2$ are {\bf not strictly isospectral}. 
\newline However, one can construct, a one parameter family of {\bf strictly isospectral} potentials $\hat{U_1}(\lambda, \rho)$, explointing the fact that for a given $U_1(\rho)$, $U_0(\rho)$ and $W(\rho)$ are not unique (see Eqs. (12) \& (13)), since the Riccati equation is a nonlinear one. Following \cite{cooper-1995, khare-1989, darboux-1982}, it can be shown that the most general superpotential satisfying Riccati equation for $U_1(\rho)$ (Eq. (16)) is given by 
\begin{equation}
\hat{W}(\rho)=W(\rho)+\frac{\hbar}{\sqrt{2m}}\frac{d}{d\rho}\log [I_{0}(\rho) +
\lambda]
\end{equation}
where $\lambda$ is a constant of integration, and $I_0$ is given by 
\begin{equation}
I_{0}(\rho)={\displaystyle\int}_{\rho^{\prime}=0}^{\rho} {[\psi_{0}(\rho^{\prime})]}^{2} d\rho^{\prime},
\end{equation}
in which $\psi_0(\rho)$ is the normalized ground state wave function of $U_0(\rho)$. The potential 
\begin{equation}
\hat{U_1}(\lambda, \rho)= \hat{W}^{2}(\rho)-\frac{\hbar}{\sqrt{2m}}
\hat{W}^{\prime}(\rho)
= U_1(\rho) - 2\frac{\hbar^{2}}{2m}
\frac{d^{2}}{d\rho^{2}}\log [I_0(\rho) + \lambda], 
\end{equation}
has the same SUSY partner $U_2(\rho)$. $\hat{U_1}(\lambda, \rho)$ has its ground state at zero energy with the corresponding wavefunction given by 
\begin{equation}
\hat{\psi_1}(\lambda, \rho)= \frac{\psi_1}{I_0+\lambda}.
\end{equation}
Hence, potentials $\hat{U_1}(\lambda, \rho)$ and $U_1(\rho)$ are {\bf strictly isospectral}. The parameter $\lambda$ is arbitrary in the intervals $-\infty<\lambda<-1$ and $0<\lambda<\infty$. $I_{0}(\rho)$ lies between 0 and 1, so the interval $-1\leq \lambda\leq 0 $ is forbidden, in order to bypass singularities in $\hat{U_1}(\lambda, \rho)$. For $\lambda \rightarrow \pm \infty$, $\hat{U_1} \rightarrow U_1$ and for $\lambda \rightarrow 0+$, $\hat{U_1}$ develops a narrow and deep attractive well in the viscinity of the origin. This well-barrier combination effectively traps the particle giving rise to a sharp resonance. This method has been 
tested successfully for 3D finite square well potential \cite{das-2001} choosing parameters capabe of supporting one or more resonance state(s) in addition to one bound state. Nuclei $^{18, 20}$C have in their ground states $T=1, J^{\pi}=0^+$ and there exists a resonance state of the same $J^{\pi}$. Thus, the forgoing procedure starting from the ground state of $^{18, 20}$C will give $T=1, J^{\pi}=0^+$ resonance(s). In an attempt to  search for the correct resonance energy, we compute the probability of finding the system within the well region of the potential $\hat{U_1}(\lambda, \rho)$ corresponding to the energy $E$ ($>0$) by integrating the probability density up to the top of the barrier:
\begin{equation}
G(E)=\int_{\rho^{\prime}=0}^{\rho_B} |\hat{\psi_E}(\rho^{\prime},\lambda)|^2d\rho^{\prime}
\end{equation}
where $\rho_B$ indicates position of the top of the barrier component of the potential $\hat{U_1}(\lambda, \rho)$ for a chosen $\lambda$. Here $\hat{\psi_E}(\lambda, \rho)$ that represents the solution of the potential $\hat{U_1}(\lambda, \rho)$, corresponding to a positive energy $E$, is normalized to have a constant amplitude in the assymptotic region. Plot of the quantity $G(E)$ against increasing $E$ ($E > 0$) shows a peak at the resonance energy $E=E_R$. Choice of $\lambda$ has to be made judiciously to avoid numerical errors entering in the wavefunction in the extremely narrow well for $\lambda\rightarrow 0+$. The width of resonance can be obtained from the mean life of the state using the energy-time uncertainty relation. The mean life is reciprocal to the decay constant. And the decay constant is the product of the number of hit per unit time on the barrier and the corresponding probability of tunneling through the barrier. 

\section{Results and discussions}
Eq.(6) is solved for the GPT n-n potential [31] and core-n SBB potential [32]. The range parameter for the core-n potential $b_{cn}$ is slightly adjusted to match the experimetal ground state spectra. The calculated two-neutron separation energies ($S_{2n}$), the relative convergence in energy (=$\frac{E(K_{max}+4)-E(K_{max})}{E(K_{max}+4)}$) and the rms matter radii (R$_A$) for gradualy increasing $K_{max}$ are listed in Table 1 for both of $^{18}$C and $^{20}$C. Although the computed results indicate a clear convergence trend with increasing $K_{max}$, it is far away from full convergence even at $K_{max}=24$. For this reason, we used an extrapolation technique succesfully used for atomic systems \cite{das-1994, khan-2012} as well as for nuclar system \cite{khan-2001}, to get the converged value of about 4.91 MeV for $^{18}$C and 3.51 MeV for $^{20}$C as shown in columns 2 and 4 of Table 2. Partial contribution of the different partial waves to the two-neutron separation enegies corresponding to $l_x = 0, 1, 2, 3, 4$ are presented in Table 3. Variation of the two-neutron separation as a function of $K_{max}$ is shown in Figure 1 for both the nuclei $^{18}$C and $^{20}$C. In Figure 2 we have shown the relative convergence trend in enrgies as a function of $K_{max}$. In Figures 3 and 4 we have presented a 3D view of the correlation density profile of the halo nuclei $^{18}$C and $^{20}$C. While in Figures 5 and 6 we have shown the 2D projection of the 3D probability density distribution. The figures clearly indicate the halo structure of the nuclei comprising a dense core surrounded by low density tail.
\newline After getting the ground state energy and wavefunctions we constructed the isospectral potential invoking principles of SSQM to investigate the resonant states. The lowest eigen potential obtained for the ground states of $^{18}$C and $^{20}$C as shown in yellow lines in Figures 7 and 8 exhibits shallow well followed by a broad and low barrier. This low well-barrier combination may indicate resonant states. However, since the well is very shallow and the barrier is not sufficiently high, the resonance width is very large and a numerical calculation of the resonant state is quite challenging. Hence, we constructed the one-parameter family of isospectral potentials $\hat{U}(\lambda, \rho)$ following Eq.(19) by appropriate selection $\lambda$ parameter values, such that a narrow and sufficiently deep well followed by a high barrier is obtained which are also shown in Figures 7 and 8 in representative cases. The enhanced well-barrier combination effectively traps the particles to form a strong resonant state. Calculated parameters of the isospectral potential for some $\lambda $ values, along with the original lowest eigen potential $U_0(\rho)$ (which corresponds to $\lambda\rightarrow\infty$) are presented in Table 4. One can, for example, note from Table 4 under $^{18}$C that, when $\lambda$ changes from = 0.1 to 0.0001, the depth of the well increases from -24.2 MeV at 2.7 fm to -247.6 MeV at 1.3 fm while the height of the barrier increases from 5.4 MeV at 5.1 fm to 121.5 MeV at 1.9 fm. The same trend is observed for $^{20}$C also. Thus the application of SSQM produces a {\bf dramatic effect} in the isospectral potential $\hat{U_1}(\lambda, \rho)$ as $\lambda$ approaches 0+. Further smaller positive values of $\lambda$ are not desirable since that will make the well too narrow to compute the wave functions accurately by a standard numerical technique. 
The probability of trapping, $G(E)$, of the particle within the enhanced well-barrier combination as a function of the particle energies E shown in Figures 9 and 10 exhibit resonance peak at the energies $E_R\simeq 1.89$ MeV for $^{18}$C and at energy $E_R\simeq 3.735$ MeV for $^{20}$C respectively. It is interesting to see that the resonance energy is independent of the $\lambda$ parameter. The enhancement of accuracy in the determination of $E_R$ is the principal advantage of using Supersymmetric formalism. Since $\hat{U_1}(\lambda;\rho)$ is strictly isospectral with $U(\rho)$, any value of $\lambda$ is admissible in principle. However, a judicious choice of $\lambda$ is necessary for accurate determination of the resonance energy. The calculated two-neutron separation energies are in excellent agreement with the observed values $4.910\pm 0.030$ MeV for $^{18}$C and $3.510\pm 0.240$ MeV for $^{20}$C \cite{audi1-2003} and also with results of Yamaguchi et al 
\cite{yamaguchi-2011} as presented in Table 5. The calculated RMS matter radii also agree fairly with the experimental values \cite{ozawa1-2001}. 

\section{Summary and conclusions}
In this communication we have investigated the structure of $^{18, 20}$C using hyperspherical harmonics expansion method assuming $^{16,18}$C$+n+n$ three-body model. Standard GPT \cite{gogny-1970} potential is chosen for the $n-n$ pair while a three-term Gaussian SBB potential \cite{sack-1954} with adustable range parameter is used to compute the ground state energy and wavefunction. The one parameter familly of isospectral potentials constructed using the ground state wavefunctions succesfully explains the resonance states in both the systems. The method is a robust one and can be applied for any weakly bound system even if the system lacks any bound ground state.

\ack
\hspace*{1cm} This work has been supported by computational facility at Aliah University, India.

\newpage
\section{Tables}   
\begin{center}
\begin{table}[h]
\caption{Two neutron separation energies, their relative convergence and rms matter radius of $^{18}$C and $^{20}$C for different $K_{max }$ in their ground states.}
\begin{tabular}{|c|c|c|c|c|c|c|}\hline\hline
$System$&\multicolumn{3}{c}{$^{18}$C ($^{16}$C+n+n)}\vline &\multicolumn{3}{c}{$^{20}$C ($^{18}$C+n+n)}\vline\\
\cline{2-4}\cline{5-7}
$K_{max}$&$S_{2n}$(MeV)&Rel. Convergence&$R_{A}$&$S_{2n}$(MeV)&Rel. Convergence&$R_{A}$\\\hline
4 &3.97843&0.10341&2.8297&2.55712&0.15771&2.9832\\\hline
8 &4.43727&0.03861&2.7913&3.03591&0.05480&2.9431\\\hline
12&4.61546&0.02521&2.7647&3.21193&0.03493&2.9145\\\hline
16&4.73481&0.01344&2.7425&3.32818&0.01944&2.8867\\\hline
20&4.79933&0.00638&2.7233&3.39416&0.00984&2.8628\\\hline
24&4.83013&       &2.7156&3.42789&       &2.8479\\\hline
\end{tabular}
\end{table}
\end{center}
\begin{center}
\begin{table}[h]
\caption{Extrapolated values of two-neutron separation energies and their relative convergences for the ground (J$^{\pi}$=0$^+$) state of $^{18}$C and $^{20}$C.}
\centering
\begin{tabular}{|c|c|c|c|c|}\hline\hline
$System$&\multicolumn{2}{c}{$^{18}$C ($^{16}$C+n+n)}\vline &\multicolumn{2}{c}{$^{20}$C ($^{18}$C+n+n)}\vline\\
\cline{1-3}\cline{4-5}
$K_{max}$&$S_{2n}(MeV)$&Rel. Convergence&$S_{2n}(MeV)$&Rel. Convergence\\\hline
24 &4.83013120&0.00456309&3.42789570&0.00669549\\\hline
28 &4.85227259&0.00303493&3.45100186&0.00446219\\\hline
32 &4.86704371&0.00209706&3.46646992&0.00308899\\\hline
36 &4.87727164&0.00149547&3.47721099&0.00220653\\\hline
40 &4.88457636&0.00109515&3.48490054&0.00161829\\\hline
44 &4.88993151&0.00082034&3.49054929&0.00121388\\\hline
48 &4.89394622&0.00062664&3.49479156&0.00092838\\\hline
52 &4.89701487&0.00048691&3.49803908&0.00072217\\\hline
56 &4.89940043&0.00038406&3.50056708&0.00057019\\\hline
60 &4.90128282&0.00030700&3.50256423&0.00045621\\\hline
64 &4.90278798&0.00024834&3.50416285&0.00036934\\\hline
68 &4.90400585&0.00020305&3.50545757&0.00030221\\\hline
72 &4.90500182&0.00016763&3.50651729&0.00024967\\\hline
76 &4.90582420&0.00013961&3.50739299&0.00020807\\\hline
80 &4.90650922&0.00011721&3.50812293&0.00017479\\\hline
84 &4.90708439&0.00009913&3.50873623&0.00014791\\\hline
88 &4.90757089&0.00008441&3.50925527&0.00012599\\\hline
92 &4.90798515&0.00007231&3.50969749&0.00010800\\\hline
96 &4.90834009&0.00006232&3.51007659&0.00009311\\\hline
100&4.90864598&0.00005399&3.51040343&0.00008069\\\hline
104&4.90891099&0.00004693&3.51068673&0.00007028\\\hline
108&4.90914137&0.00004119&3.51093349&0.00006149\\\hline
112&4.90934358&0.00003612&3.51114943&0.00005405\\\hline
116&4.90952089&0.00003186&3.51133920&0.00004769\\\hline
120&4.90967732&0.00002821&3.51150666&0.00004224\\\hline
124&4.90981582&0.00002507&3.51165499&0.00003755\\\hline
128&4.90993891&0.00002235&3.51178684&0.00003349\\\hline
132&4.91004866&0.00001999&3.51190444&0.00002996\\\hline
136&4.91014683&0.00001794&3.51200967&0.00002689\\\hline
140&4.91023492&0.00001614&3.51210410&0.00002419\\\hline
144&4.91031418&0.00001456&3.51218909&0.00002184\\\hline
148&4.91038569&0.00001318&3.51226579&0.00001976\\\hline
152&4.91045039&0.00001195&3.51233519&0.00001792\\\hline
156&4.91050905&0.00001086&3.51239814&0.00001629\\\hline
160&4.91056237&0.00000989&3.51245537&0.00001484\\\hline
164&4.91061094&0.00000903&3.51250750&0.00001355\\\hline
168&4.91065528&0.00000826&3.51255511&0.00001239\\\hline
172&4.91069584&0.00000757&3.51259866&0.00001137\\\hline
176&4.91073301&0.00000695&3.51263858&0.00001044\\\hline
180&4.91076715&0.00000639&3.51267525&0.00000960\\\hline
184&4.91079855&0.00000589&3.51270898&0.00000885\\\hline
188&4.91082748&0.00000544&3.51274007&0.00000817\\\hline
192&4.91085418&0.00000503&3.51276876&0.00000755\\\hline
196&4.91087887&0.00000465&3.51279529&0.00000699\\\hline
200&4.91090172&          &3.51281985&          \\\hline
........&........&........&........&........\\\hline
$\infty$&4.91124921&&3.51319416&\\\hline
\end{tabular}
\end{table}
\end{center}
\begin{center}
\begin{table}[h]
\caption{Partial contribution of different $l_x$ partial waves to the two-neutron separation energy in the ground states of $^{18}$C and $^{20}$C corresponding to different $K_{max }$.}
\centering
\begin{tabular}{|c|c|c|c|c|c|c|c|c|c|c|}\hline\hline
$System$&\multicolumn{5}{c}{$^{18}$C}\vline &\multicolumn{5}{c}{$^{20}$C}\vline\\
\cline{1-6}\cline{7-11}
$K_{max}$&\multicolumn{5}{c}{$E_{lx}$for $l_x$=}\vline	&\multicolumn{5}{c}{$E_{l_x}$ for $l_x$=}\vline\\
\cline{2-11}
  &0       &1      &2      &3      &4       &0 &1&2&3&4\\\hline
4 &2.886 &0.068&1.149&0.000&0.000&2.471&0.235&0.036&0.000&0.000\\\hline
8 &3.283 &0.072&1.158&0.001&0.100&2.902&0.232&0.019&1.034&0.001\\\hline
12&3.414 &0.078&1.158&0.001&0.098&3.073&0.231&0.014&1.036&0.001\\\hline
16&3.487 &0.079&1.166&0.001&0.098&3.1948&0.2278&0.0136&1.040&0.001\\\hline
20&3.536 &0.081&1.181&0.001&0.096&3.262&0.225&0.013&1.049&0.001\\\hline
24&3.566 &0.082&1.199&0.001&0.094&3.295&0.225&0.013&1.080&0.001\\\hline
\end{tabular}
\end{table}
\end{center}
\begin{center}
\begin{table}[h]
\caption{Data to show the effects of the parameter $\lambda$ on the depth of the well and height of the barrier, in the isospectral potential constructed from the lowest eigen potential and the corresponding wavefunction for $^{18}$C and $^{20}$C.}
\begin{tabular}{|c|c|l|c|c|c|c|c|c|}\hline\hline
$System$&\multicolumn{4}{c}{$^{18}$C}\vline &\multicolumn{4}{c}{$^{20}$C}\vline\\
\cline{1-5}\cline{6-9}
$\lambda$&\multicolumn{2}{c}{Potential	Well}\vline	&\multicolumn{2}{c}{Potential	Barrier}\vline	&\multicolumn{2}{c}{Potential	Well}\vline	&\multicolumn{2}{c}{Potential	Barrier}\vline	\\
\cline{2-9}
&$Depth, V_0 $&At, $r$&$Height, V_B$&At, $r$&$Depth, V_0$&At, $r$&$Height, V_B$&At, $r$\\
&$(MeV)$&$(fm)$&$(MeV)$&$(fm)$&$(MeV)$&$(fm)$&$(MeV)$&$(fm)$\\
\hline
100000 &-9.301  &3.092&2.702  &20.099&-11.076&3.069&3.085&7.520\\\hline
100    &-9.311  &3.092&2.712  &20.099&-11.086&3.069&3.094&7.514\\\hline
50     &-9.337  &3.091&2.712  &20.099&-11.163&3.065&3.106&7.504\\\hline
1      &-11.590 &3.019&2.719  &20.095&-17.294&2.824&3.938&6.708\\\hline
0.1    &-24.220 &2.664&5.394  &5.088&-41.931&2.320&10.924&3.983\\\hline
0.01   &-62.846 &2.109&19.797 &3.424&-95.444&1.811&36.751&2.840\\\hline
0.001  &-136.144&1.637&56.596 &2.521&-154.453&1.389&84.198&2.173\\\hline
0.0001 &-247.602&1.274&121.482&1.937&-270.962&1.000&141.518&1.731\\\hline
0.00001&-384.056&0.986&218.025&1.520&-479.005&0.724&241.024&1.218\\\hline
\end{tabular}
\end{table}
\end{center}
\begin{table}[h]
\begin{center}
\caption{\label{tab:table3} Comparison of the calculated data with those found in the literature for $^{18}$C and $^{20}$C halo nuclei.}
\begin{tabular}{|c|c|c|c|r|}\hline
Nuclide&State& Observables& Present work & Others work\\\hline
$^{18}$C&$0^{+}$             & BE      &4.9064 MeV&4.910 $\pm$0.030 MeV\cite{audi1-2003}\\
&&&&4.91 MeV\cite{yamaguchi-2011}\\
&             & $R_A$      & 2.7156 fm&2.82$\pm0.04$ fm\cite{ozawa1-2001}\\
&0$^{+}_{1}$&$E_{R}$ &1.89 MeV&-\\\hline  
$^{20}$C&$0^{+}$             & BE      &3.5065 MeV&3.510 $\pm$0.240 MeV\cite{audi1-2003}\\
&&&&3.51 MeV\cite{yamaguchi-2011}\\
&             & $R_A$      &2.8479 fm&2.98$\pm0.05$ fm \cite{ozawa1-2001}\\
&0$^{+}_{1}$&$E_{R}$ &3.735 MeV&-\\\hline                             
\end{tabular}
\end{center}
\end{table}

\section{Graphs and Figures}
\begin{figure}[h]
\begin{minipage}{20pc}
\includegraphics[width=20pc,height=15pc]{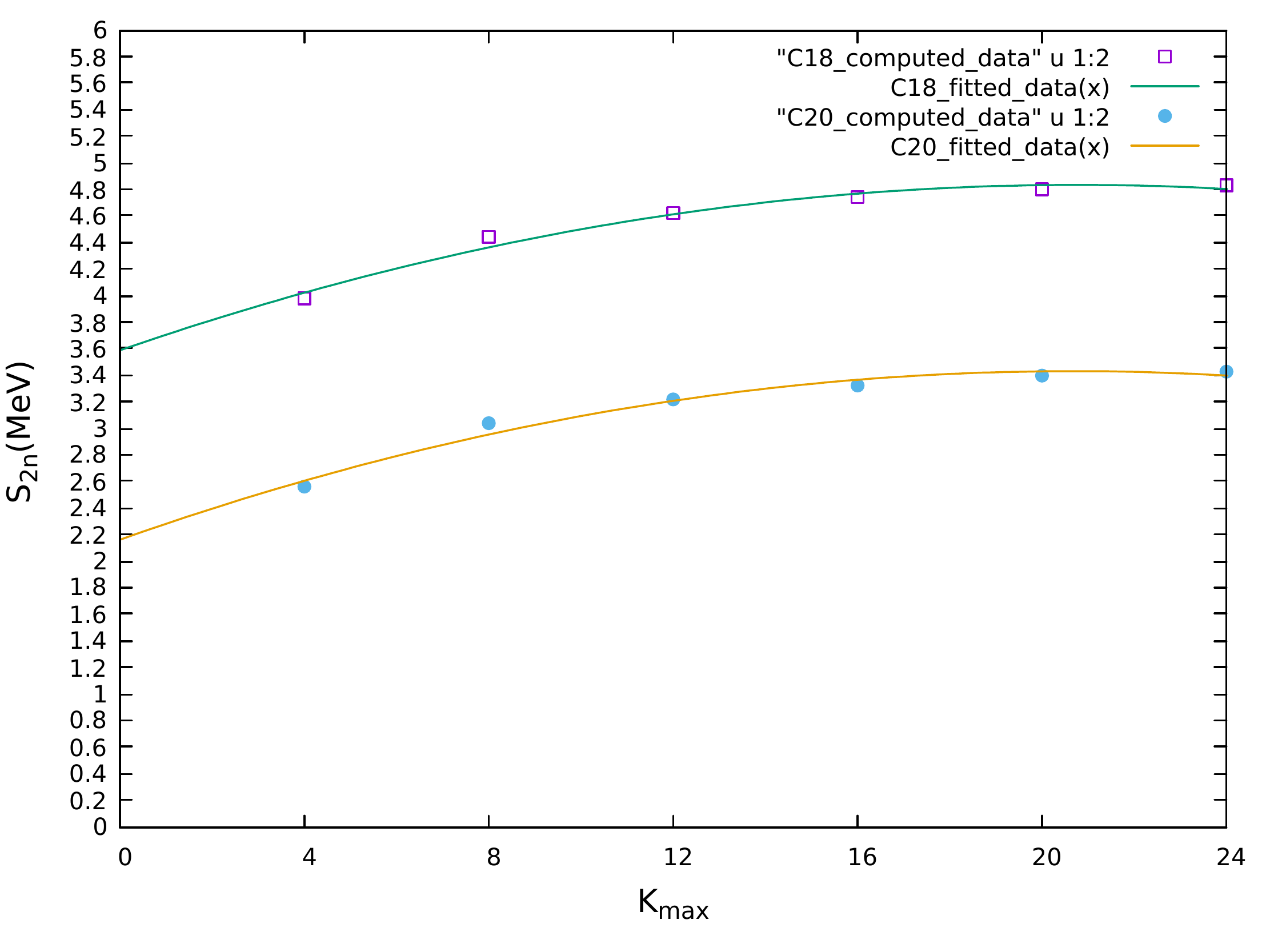}
\caption{Plot of two-neutron separation energy (S$_{2n}$) as a function of K$_{max}$ for $^{18}$ (upper curve) and $^{20}$C (lower curve).}
\end{minipage}\hspace{2pc}
\begin{minipage}{18pc}
\includegraphics[width=20pc,height=15pc]{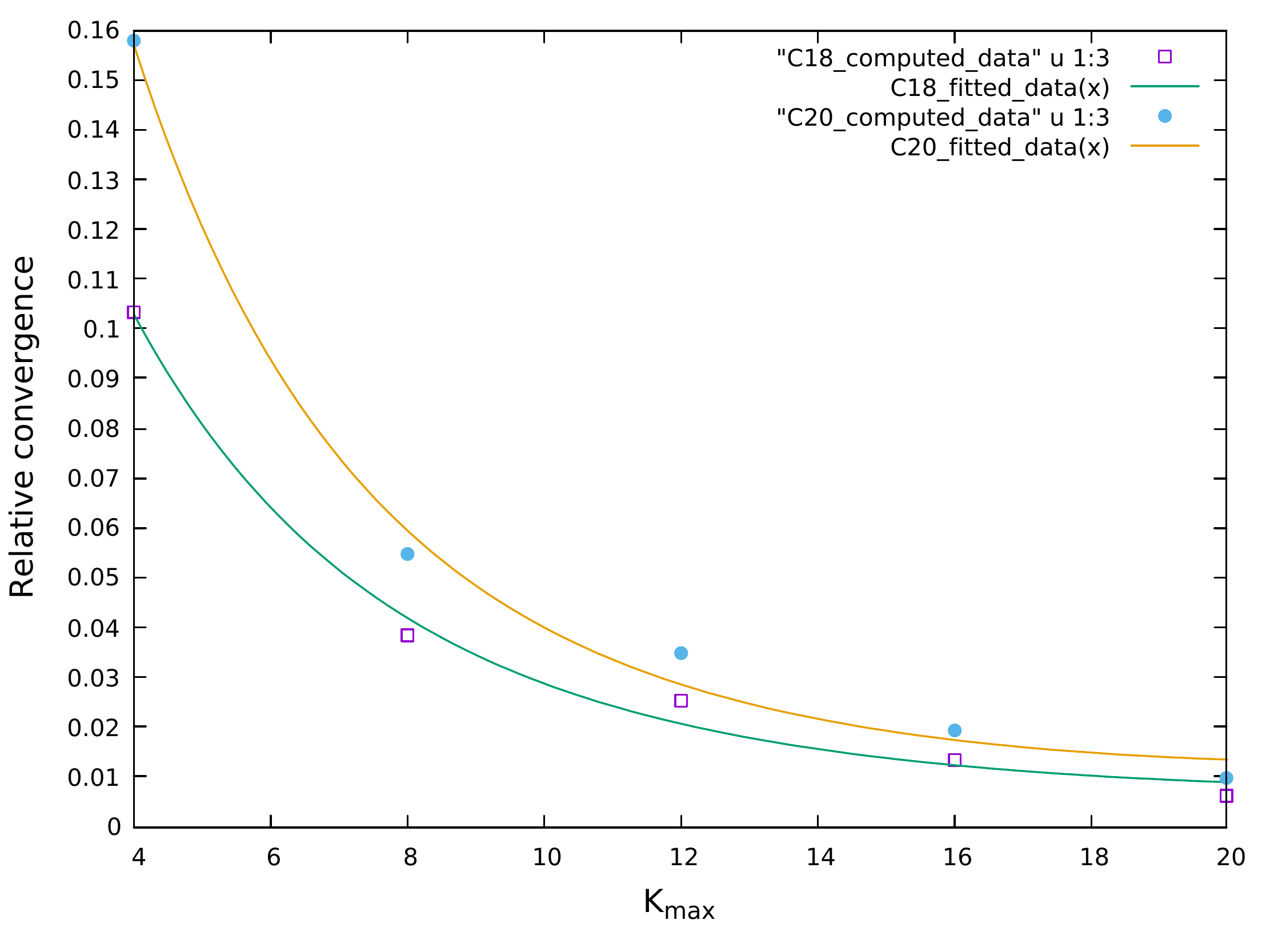}
\caption{Plot of relative convergance rate with respect to increasing K$_{max}$ for $^{18}$ (lower curve) and $^{20}$C (upper curve).}
\end{minipage}
\end{figure}
\begin{figure}[h]
\begin{minipage}{20pc}
\includegraphics[width=20pc,height=15pc]{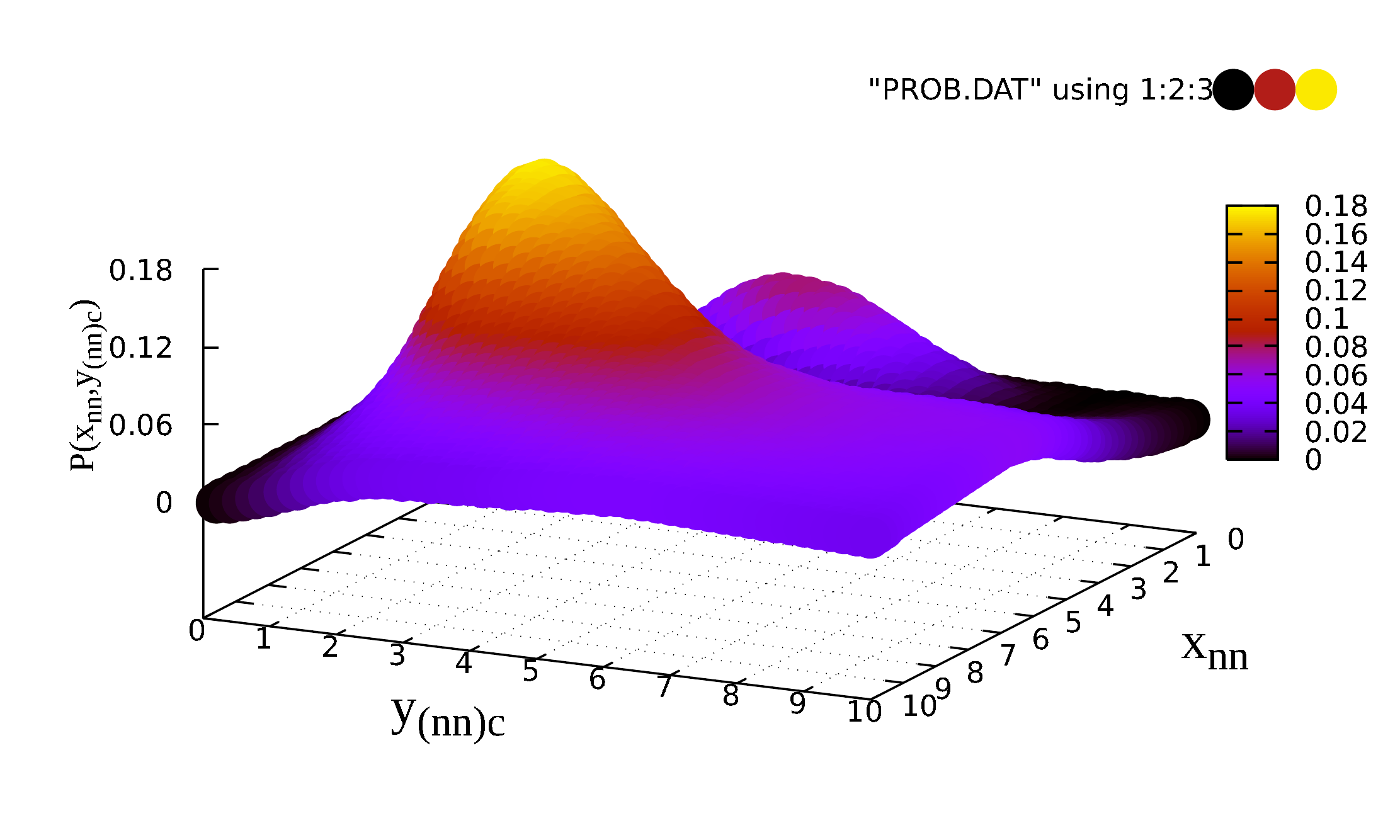}
\caption{Correlation density plot for the ground state (0$^{+}$) state of 2n-halo $^{18}$C nucleus as a function of the Jacobi coordiantes $x_{nn}$ and $y_{(nn)c}$.}
\end{minipage}\hspace{2pc}
\begin{minipage}{18pc}
\includegraphics[width=20pc,height=15pc]{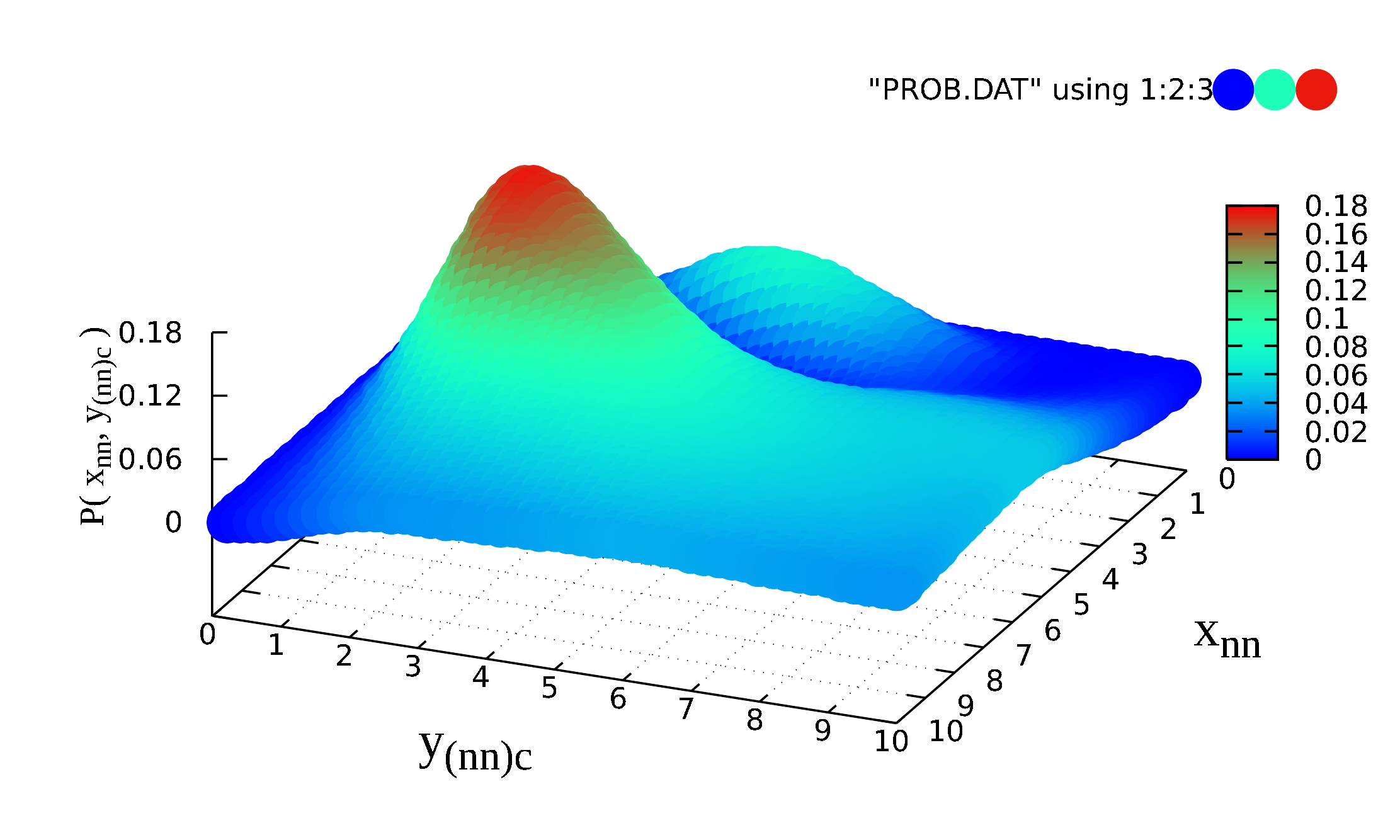}
\caption{Correlation density plot for the ground state (0$^{+}$) state of 2n-halo $^{20}$C nucleus as a function of the Jacobi coordiantes $x_{nn}$ and $y_{(nn)c}$.}
\end{minipage}
\end{figure}
\begin{figure}[h]
\begin{minipage}{20pc}
\includegraphics[width=20pc,height=15pc]{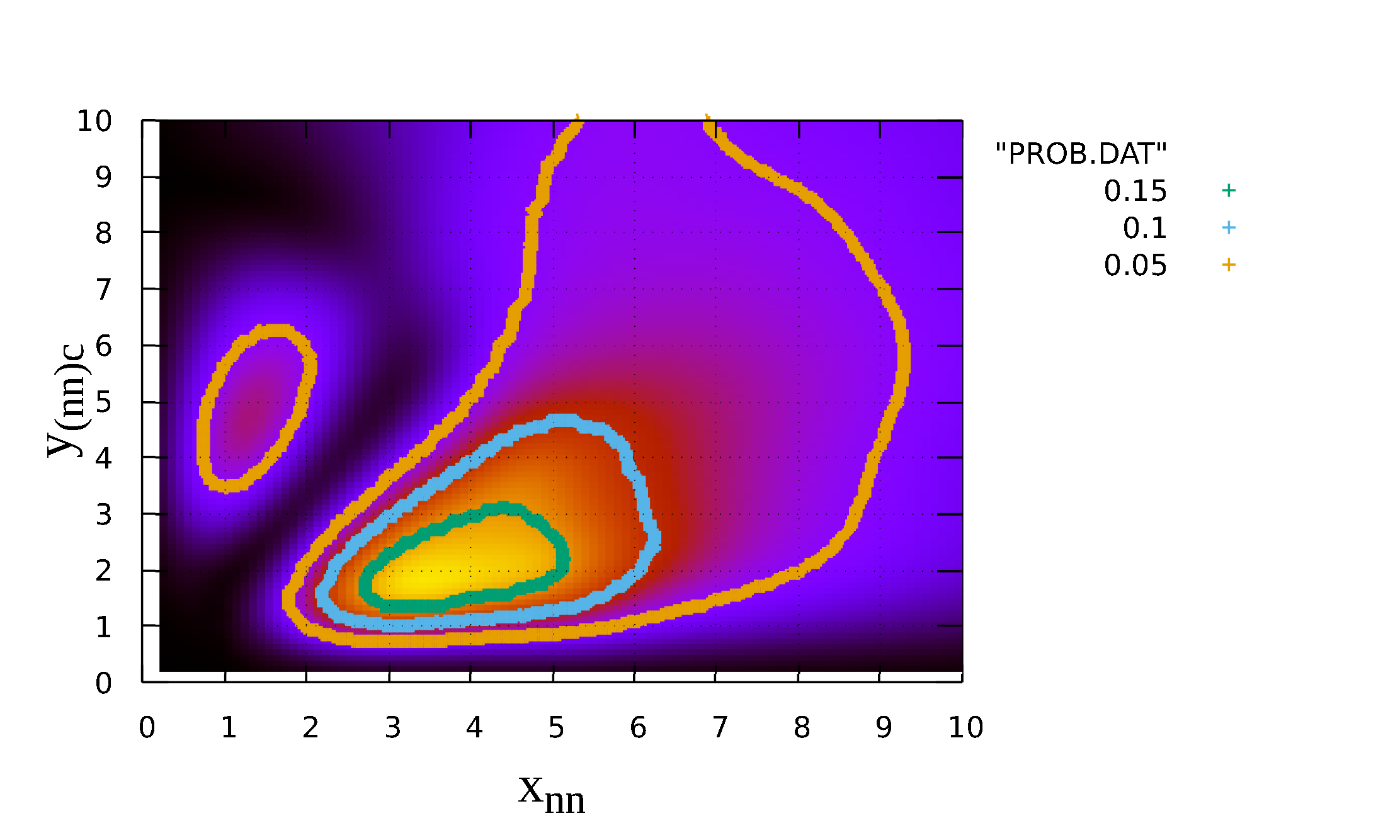}
\caption{2D projection of the correlation density plot for the0$^{+}$ state of 2n-halo $^{18}$C nucleus as a function of the Jacobi coordiantes $x_{nn}$ and $y_{(nn)c}$.}
\end{minipage}\hspace{2pc}
\begin{minipage}{18pc}
\includegraphics[width=20pc,height=15pc]{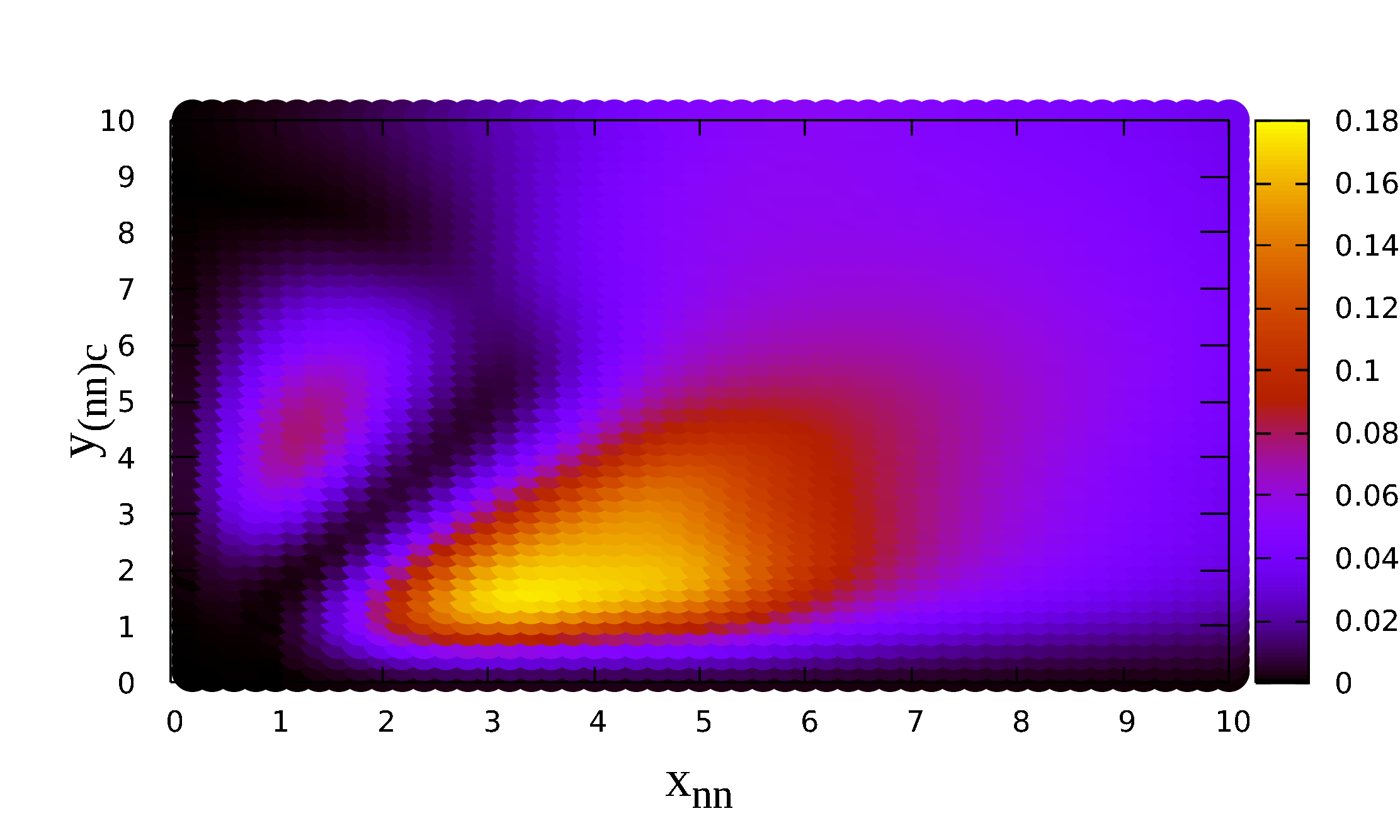}
\caption{2D projection of the correlation density plot for the0$^{+}$ state of 2n-halo $^{20}$C nucleus as a function of the Jacobi coordiantes $x_{nn}$ and $y_{(nn)c}$}
\end{minipage}
\end{figure}

\begin{figure}[h]
\begin{minipage}{20pc}
\includegraphics[width=20pc,height=15pc]{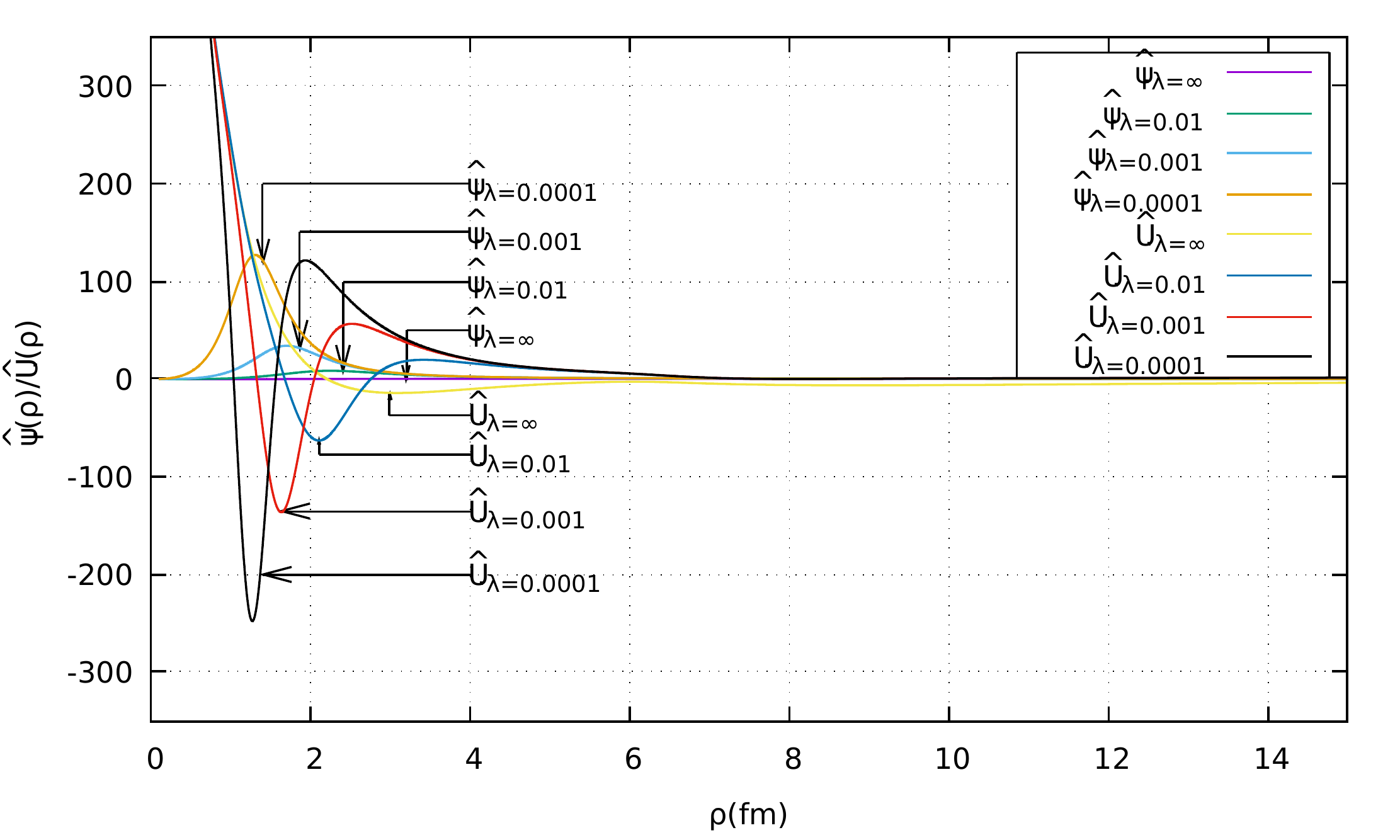}
\caption{Graphical presentation of potentials and corresponding wavefunctions of $^{18}C$ for few representative values of $\lambda$ ($=\infty$(original potential), 0.01, 0.001 and 0.0001.}
\end{minipage}\hspace{2pc}
\begin{minipage}{18pc}
\includegraphics[width=20pc,height=15pc]{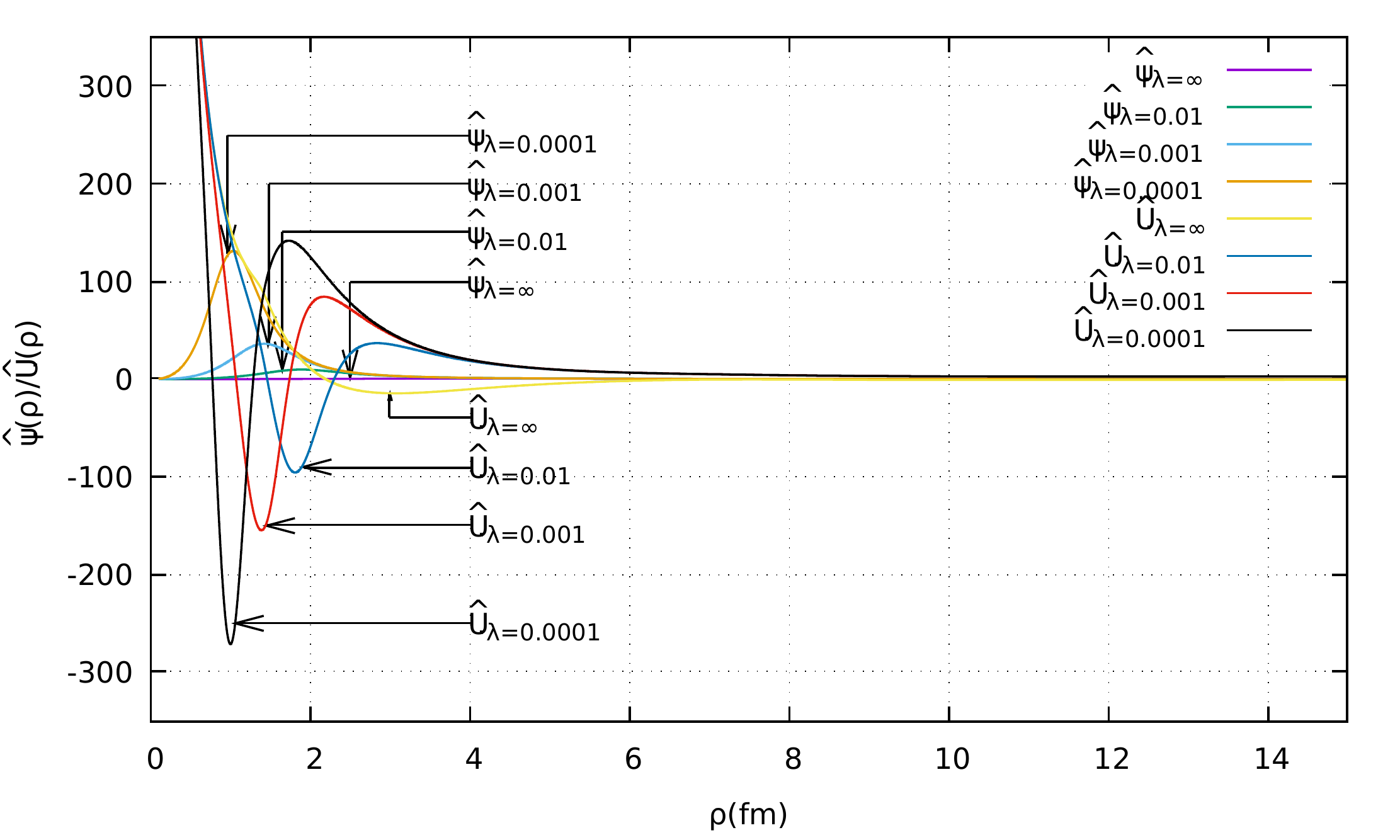}
\caption{Graphical presentation of potentials and corresponding wavefunctions of $^{20}C$ for few representative values of $\lambda$ ($=\infty$(original potential), 0.01, 0.001 and 0.0001.}
\end{minipage}
\end{figure}
%
\begin{figure}[h]
\begin{minipage}{20pc}
\includegraphics[width=20pc,height=15pc]{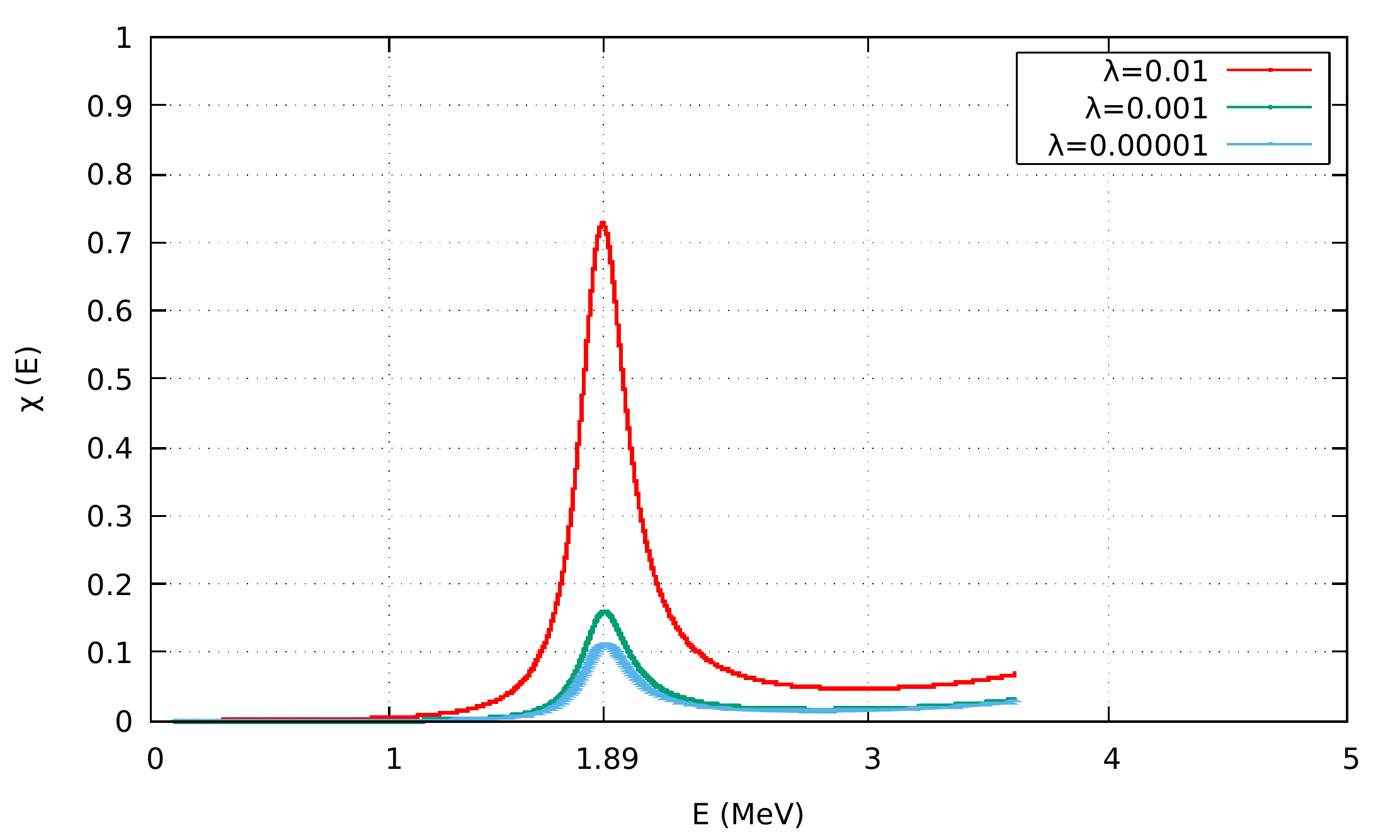}
\caption{Plot of $G(E)$ as a function of energy demostrating the resonant state(s) in $^{18}C$.}
\end{minipage}\hspace{2pc}
\begin{minipage}{18pc}
\includegraphics[width=20pc,height=15pc]{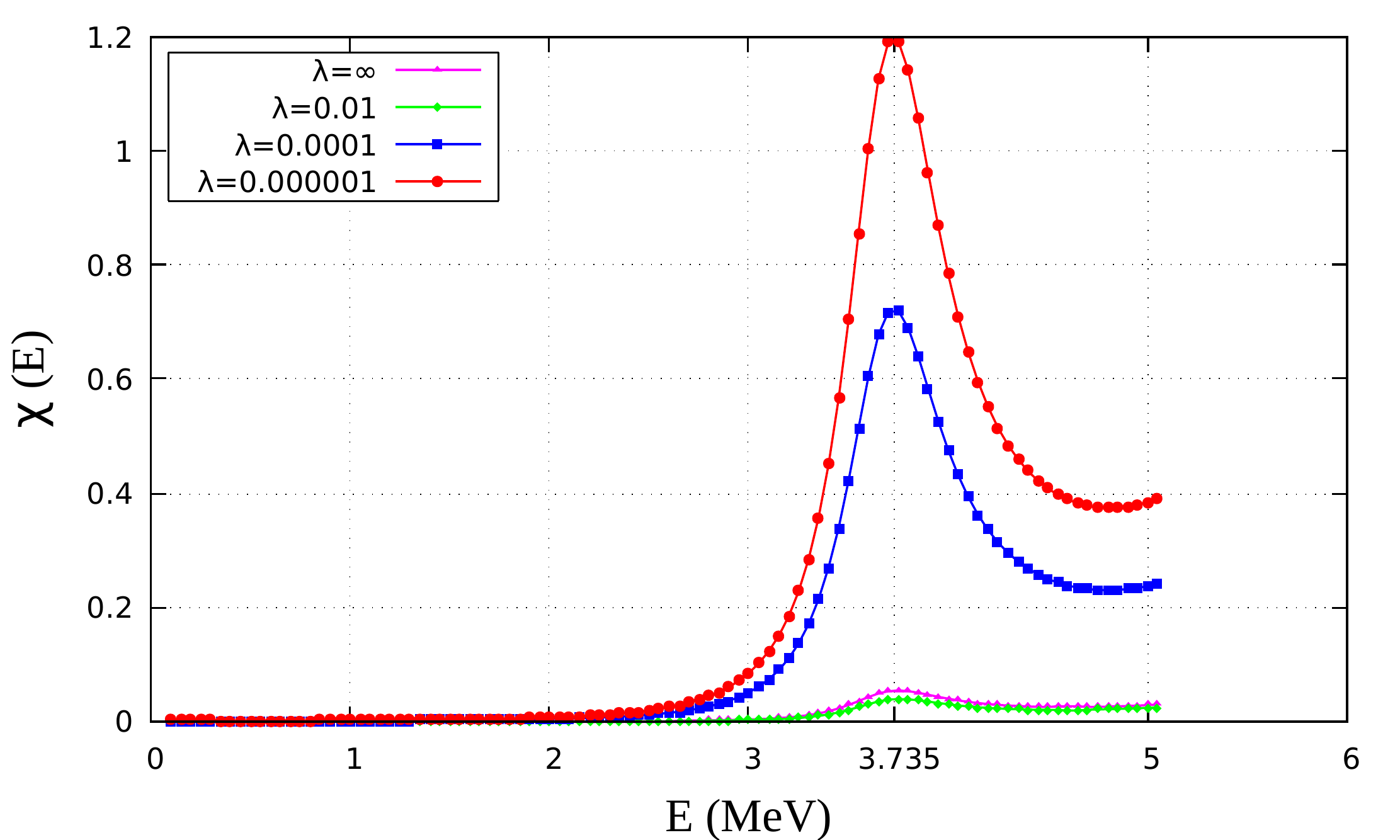}
\caption{lot of $G(E)$ as a function of energy demostrating the resonant state(s) in $^{20}C$.}
\end{minipage}
\end{figure}

\section*{References}
\begin{thebibliography}{9}
\bibitem{tanihata-1985} I. Tanihata et al 1985 Phys. Rev. Lett. {\bf 55} N0.24, 2676 
\bibitem{kobayashi-2012} N. Kobayashi {\it et al} 2012 {\it Phys. Rev. C.} {\bf 86} 054604
\bibitem{hwash-2017} W. S. Hwash 2017 {\it Turk. J. Phys.} {\bf 41} 151-159
\bibitem{jensen-2000} A. S. Jensen and K. Riisager 2000 Phys. Lett. {\bf B480}, No. 1, 39-44
\bibitem{schwab-2000} W. Schwab et al 2000 Z. Phys. A350 (1995) 283
\bibitem{tanaka-2010} K. Tanaka {\it et al.} 2010 {\it Phys. Rev. Lett.} {\bf 104} 062701
\bibitem{audi-2003} G. Audi et al 2003 Nucl. Phys. A {\bf 729}, No.1, 3-128
\bibitem{acharya-2013} B. Acharya, C. Ji, D.R. Phillips 2013 {\it Phys. Lett. B.} {\bf 723} 196
\bibitem{gaudefroy-2012} Gaudefroy, L. et al. Direct mass measurements of $^{19}$B, $^{22}$C, $^{29}$F, $^{31}$Ne, $^{34}$Na and other light exotic nuclei. Phys. Rev. Lett. 109, 202503 (2012).
\bibitem{togano-2016} Togano, Y. et al. Interaction cross section study of the two-neutron halo nucleus $^{22}$C. Physics Letters B 761, 412 – 418 (2016).
\bibitem{saaf-2014} Daniel S\"{a}\"{a}f and Christian Forssen (2014) Phys. Rev. {\bf C89}, No.1, 011303(R)
\bibitem{nesterov-2010} A. V. Nesterov et al 2010 {\it Phys. of Part. and Nuclei} {\bf 41} No.5, 716-765 
\bibitem{korennov-2004} S. Korennov and Pierre Descouvemont 2004 Nucl. Phys. A740, No.3, 249-267
\bibitem{cobis-1997} A. Cobis, D. V. Federov and A. S. Jensen 1997 Phys. Rev. Lett.
{\bf 79}, 2411; \\ A. Cobis, D.V. Federov and A.S. Jensen 1998 Phys. Rev. C {\bf 58}, 1403 
\bibitem{csoto-1993} A. Cs\'{o}t\'{o} 1993 Phys. Lett. {\bf B315}. 24; 1993 Phys. Rev. {\bf C48} 165; A. Cs\'{o}t\'{o} 1993 {\it ibid.} {\bf C48} 165; A. Cs\'{o}t\'{o} 1994 {\it ibid.} {\bf C49} 3035
\bibitem{ayoma-1995} S. Aoyama, S. Mukai, K. Kato and K. Ikeda 1995 Prog. Theor. Phys. {\bf 94} 343
\bibitem{tanaka-1997} N. Tanaka, Y. Suzuki, K. Varga 1997 Phys. Rev. C56, 562 
\bibitem{vasilevsky-2001} V. Vasilevsky, A. V. Nesterov, F. Arickx, J.Broeckhove 2001 Phys. Rev. C63, 034607 
\bibitem{ogata-2013} Kazuyuki Ogata, Takayuki Myo, Takenori Furumoto, Takuma Matsumoto, and Masanobu Yahiro 2013 {\it Phys. Rev. C.} {\bf 88} 024616
\bibitem{danilin-1997} Danilin, B. V., Rogde, T., Ershov, S. N., Heiberg-Andersen, H., Vaagen, J. S., Thompson, I. J., Zhukov, M. V. 1997 Phys. Rev. C55, R577 
\bibitem{ballot-1982} M. Fabre de la Ripelle et al 1982 Ann. Phys. {\bf 138}, 275-318; T. K. Das, H. T. Coelho, and M. Fabre de la Ripelle 1982 Phys. Rev. C {\bf 26}, 2288 
\bibitem{schneider-1972} T. R. Schneider 1972 Phys. Letts. B {\bf 40}, 439 
\bibitem{ballot1-1982} J. L. Ballot, M. Fabre de la Ripelle, and J.S. Levinger 1982 Phys. Rev. 
C {\bf 26}, 2301 
\bibitem{das-1982} T. K. Das, H. T. Coelho and M. Fabre de la Ripelle
1982 Phys Rev. C {\bf 26}, 2281 
\bibitem{cooper-1995} F. Cooper, A. Khare and U. Sukhatame 1995 Phys. Rep. {\bf 251}, 267
\bibitem{khare-1989} A. Khare, U. Sukhatme 1989 J. Phys. A {\bf 22}, 2847 
\bibitem{nieto-1984} M. M. Nieto 1984 Phys. Lett. {\bf B 145},  208
\bibitem{darboux-1982} G. Darboux 1882 C. R. Acad. Sci. Paris {\bf 94}, 1456 
\bibitem{das-2001} T. K. Das and B. Chakrabarti 2001 Phys. Letts. A {\bf 288}, 4 
\bibitem{gogny-1970} D. Gogny, P. Pires and R. de Tourreil 1970 Phys. Letts. 
{\bf 32B}, 591 
\bibitem{sack-1954} S. Sack, L. C. Biedenharn and G. Breit 1954 Phys. Rev. 
{\bf 93}, 321 
\bibitem{das-1994} T. K. Das, R. Chattopadhyay, and P.K. Mukherjee 1994 Phys. Rev. A
{\bf 50}, 3521 
\bibitem{khan-2012} Md. A. Khan 2012 {\it  Eur. Phys. Jour.D } {\bf 66} 83
\bibitem{khan-2001} Md. A. Khan and T. K. Das 2001 Pramana- J. Phys. {\bf 57}, 701
\bibitem{audi1-2003} G. Audi et al., 2003, Nucl. Phys. A {\bf 729}, 337
\bibitem{yamaguchi-2011} T. Yamaguchi, K. Tanaka et al, 2011, Nucl.Phys. A, {\bf 864}, 1
\bibitem{ozawa1-2001} A. Ozawa et al., 2001, Nucl. Phys. A {\bf 691}, 599
\end{thebibliography}
\end{document}